\documentclass[runningheads]{llncs}

\usepackage[T1]{fontenc}
\usepackage{graphicx}
\usepackage{amsmath}
\usepackage{microtype}
\usepackage{url}

\begin{document}

\title{On Evaluating and Improving Conversational Agents in Production}

\titlerunning{On Evaluating and Improving Conversational Agents in Production}

\author{Kasra Hosseini\inst{1} \and Wen-Sen Cheng\inst{1} \and Marco-Andrea Buchmann\inst{2} \and Emir Mulabegovic\inst{1} \and Weiwei Cheng\inst{1}}
\authorrunning{K. Hosseini et al.}
\institute{Zalando SE, Berlin, Germany \and Zalando, Switzerland}

\maketitle

\begin{abstract}
We present a framework for evaluating and improving a large-scale, multi-agent shopping assistant in production, and report lessons from its use. Offline evaluation of such a system faces three obstacles. (i) A logged conversation cannot be replayed against a modified system, because a different response changes every turn that follows. (ii) The unchanged system itself varies from run to run. Its LLM components are stochastic, and in product search the available products, their prices, and the customer's personalization signals change. (iii) Aggregate quality scores combine distinct behaviors, so they show that quality has changed but not which behavior caused the change. Our framework addresses each obstacle in turn. For a reported behavior, an \emph{Evaluation Harness} generates targeted assertions and a fixed cohort of customer scenarios. It then reproduces the behavior in a local instance of the assistant through grounded user simulation. Instead of replaying the log, the simulator writes new customer turns conditioned on the recorded messages and context. Repeated runs of the unchanged system form a stored baseline. An \emph{Improvement Orchestrator} turns the assertion results into hypotheses, implements each as an isolated modification, and compares it with the baseline using paired percentile bootstrap intervals over scenario-level differences. When an investigation ends, the harness may propose revisions to future evaluations, subject to human approval and without altering past decisions. We report production investigations with this framework. Assertion profiles showed which positions of a product carousel a failure affected, and repeated runs distinguished a real improvement from run-to-run fluctuation. Audits of the evaluation itself found a judge that lacked the evidence it needed and a model setting that was configured but not applied.
\keywords{Agentic information retrieval \and Conversational search \and Grounded user simulation \and Offline evaluation \and LLM-as-a-judge}
\end{abstract}

\section{Introduction}
\label{sec:intro}

\begin{figure}[!t]
  \centering
  \includegraphics[width=1.0\linewidth]{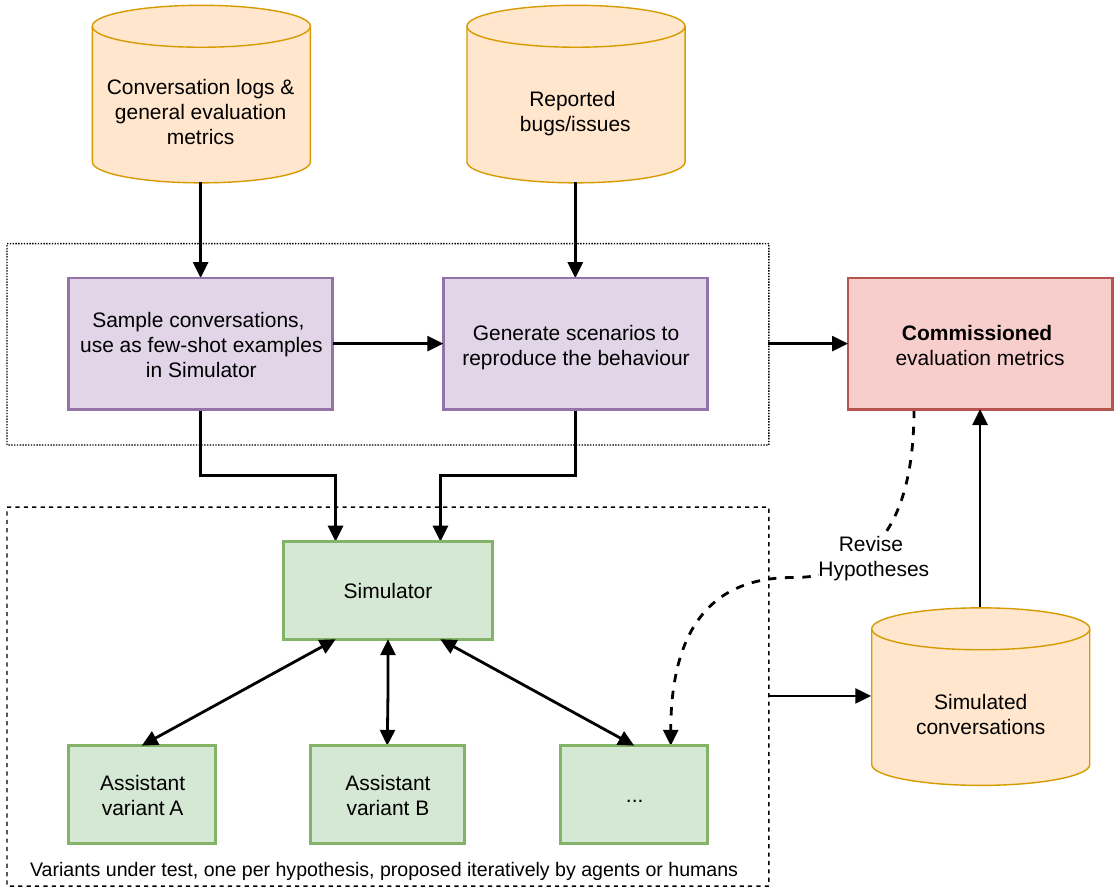}
  \caption{The proposed evaluation and improvement framework. Reported problems and low-quality production interactions are turned into scenarios that reproduce the behavior in a local version of the \emph{Production Assistant}. The simulator runs these scenarios against the unchanged system and against one modified version per hypothesis, and the resulting sessions are scored with the frozen assertions, shown in the figure as commissioned evaluation metrics. Assertion results feed back into hypothesis revision. Not shown are the stored baseline formed by repeated runs of the unchanged system, the human review of generated assertions before they are frozen, and the human-approved revision of the \emph{Evaluation Harness} between investigations (Sects.~\ref{sec:commission} and~\ref{sec:loops}).}
  \label{fig:commission}
\end{figure}

Improving production conversational agents requires evaluating whole conversations rather than the individual turns within them. Whole conversations, however, cannot be taken from production logs. A log records the conversation as the deployed system produced it. A modified system may answer differently at some turn, and the customer's later messages in the log are then no longer valid. Two further difficulties arise in production. The agents are LLM-based and therefore stochastic. In product search, the available products, their prices, and the customer's personalization signals also change between runs, so the same request can return different products. Aggregate measures such as relevance and intent understanding are useful for monitoring overall quality, but they combine distinct behaviors and do not indicate what failed or how to correct it.

Cranfield-style evaluation \cite{cleverdon1967cranfield} compares systems on a fixed test collection of topics and relevance judgments. In a conversation, later requests depend on the turns that came before. A modified system produces turns that no fixed collection has judged \cite{lipani2021how}. Interactive benchmarks and user simulators address this by letting each system generate its own conversation \cite{balog2023user,zhang2020evaluating,yao2024tau,lu2025toolsandbox}. A production decision needs more than that. It needs a precise definition of the behavior under evaluation, realistic customer interaction, and enough repeated measurements to separate a change from the run-to-run variation of a stochastic system.

We propose the framework shown in Fig.~\ref{fig:commission}. An \emph{Evaluation Harness} commissions a targeted evaluation for one behavior at a time, a grounded simulator plays the customer, an \emph{Improvement Orchestrator} proposes and tests modifications, and the harness itself is revised between investigations under human approval. We apply the framework to a multi-agent conversational shopping assistant that serves more than twenty European markets and handles millions of customer conversations each month. The framework has guided several production investigations of this assistant, and changes to the deployed system have resulted from them.

Two properties of LLM-integrated applications in production shaped the framework. The first is that we evaluate a deployed system, not a model. A guard, a router, several agents, retrieval, ranking, and personalization act on every turn (Sect.~\ref{sec:system}). The effect of changing one of them can only be seen in the output of the whole system. In our framework, the evaluation runs against a complete local instance of the assistant. A result is confirmed offline before any online test. The second is scale. The assistant handles millions of conversations a month, and the available products differ between markets and change daily. Individual conversations cannot be reviewed by people, and a single run of any comparison is dominated by chance. Problems are identified by automated assessment, cohorts are drawn from large pools by recorded sampling rules, and every comparison is repeated until the variation of the system itself is accounted for. The sections that follow describe each of these steps.

An investigation begins with a reported problem or with low-quality production interactions identified by automated quality assessments. These run daily over a sample of production conversations and produce scores, such as recommendation quality and customer satisfaction, and summaries from which personal data has been removed. The assessments indicate where to investigate but not which behavior should change. A human or the harness itself examines the report or the flagged cases, identifies the behavior to investigate, and writes a short description of it. The harness can also rank candidate behaviors into a priority list for a human to choose from.

From the description of the behavior, the \emph{Evaluation Harness} produces two artifacts. (1) Targeted assertions state, as pass or fail conditions, what the system should and should not do. (2) A fixed set of customer scenarios, the cohort, provides conversations in which that behavior can be observed. Each scenario is grounded in a relevant production conversation when one exists, and otherwise constructed from a description of what the customer wants to achieve. Once the behavior has been reproduced in a local version of the \emph{Production Assistant}, the cohort is frozen together with the assertions and the evidence fields used to judge them. Repeated runs of the unchanged local system then establish a stored baseline that captures its behavior and its run-to-run variation, so every modified version is judged against the same definition of the problem. We refer to one simulated interaction with one system version as a session and to one pass over the cohort as a run.

The simulator does not replay a logged conversation beyond its first customer message. It generates each following customer turn in response to the system version under test. Each turn is conditioned on the recorded customer messages and the customer's context, which can include personalization signals such as preferred sizes and brands. The \emph{Improvement Orchestrator} uses the detailed assertion results, not an aggregate score, to form hypotheses about which parts of the system should change. In a multi-agent system, attributing a failure to the component that caused it is difficult \cite{cemri2025why,zhang2025which}. Each hypothesis becomes an isolated modified version, compared with the stored baseline through repeated simulation and paired bootstrap intervals. A version that improves the target behavior but clearly regresses another targeted behavior or guardrail is not advanced. A favorable result is evaluated a second time, which guards against an improvement that arose by chance, and a version that passes both evaluations is presented for human review and a possible online A/B test. After an investigation, the harness can use what was learned to revise how future evaluations are constructed. Such revisions require human approval, apply only to future investigations, and do not alter the recorded decisions of the investigation that produced them.

\paragraph{Contributions.} We make three contributions. First, we propose a framework for evaluating and improving stochastic conversational agents under production conditions through commissioned evaluation, grounded user simulation, and hypothesis-driven system modification. Second, we generate behavior-specific assertions that support detailed diagnosis and guide hypothesis generation, and pair them with repeated comparisons that separate system changes from run-to-run variation. Third, we describe a human-gated self-revising harness that uses completed investigations to inform future ones while the decisions of completed investigations stay as recorded. In our production investigations, the assertion profiles located a failure at particular positions of the product carousel, and audits of the evaluation itself found defects that could have changed an improvement decision.

\section{Related Work}
\label{sec:related}

Most system comparisons in information retrieval use reusable test collections, whose limitations in assessor variation, topic coverage, assessment effort, and statistical interpretation are well documented \cite{voorhees2000variations,sanderson2005information,fuhr2017some}. Conversational search adds a further dependency, because later requests and their relevance are shaped by the preceding interaction \cite{dalton2020cast,lipani2021how}. Assessors who observe a conversation still agree fairly well with the users themselves \cite{fu2022cranfield}, so the difficulty lies in the fixed collection rather than in the judgments. User simulation addresses this difficulty by producing a new conversation for each system. It has been used in conversational recommendation, query refinement, and feedback generation \cite{balog2023user,zhang2020evaluating,salle2021studying,owoicho2023exploiting}. Recent work evaluates simulators against human behavior and changes in system quality \cite{dou2025simulatorarena,meshi2026convapparel,zhu2026realusersim}, and shows that the choice of simulator model can shift measured agent performance and that leakage from conversational history can inflate simulated results \cite{seshadri2026lostinsimulation,zhu2024reliablesimulator}. Closer to our cohort construction, Raju et al. assemble domain-specific evaluation sets for LLM judges by clustering prompts and sampling across the clusters \cite{raju2024domainsets}. They aim at a single evaluation set that covers a domain. We instead build a separate cohort and assertion set for each behavior under investigation, ground the simulated customer in the production conversations that exhibited it, and compare modified system versions on that cohort with paired repeated runs.

Our framework relies on LLM-based relevance and quality assessment both to find low-quality interactions in live traffic and to compare system versions \cite{faggioli2023perspectives,thomas2024large}. Prior work documents systematic differences between LLM and human assessments, including position and order biases, preferences for style over substance, self-preference, prompt sensitivity, variation across repeated judgments, and distinct error distributions in multimodal product relevance evaluation \cite{hosseini2024retrieve,zheng2023judging,wang2023large,wu2023style,panickssery2024selfpreference,bavaresco2024llms,mizrahi2024state,yagubyan2026coin}. When two systems perform similarly, a judge's ranking of them agrees far less often with actual task outcomes than when the gap between them is wide \cite{bodhwani2026gauge}. The behavior-specific evaluation at the center of our framework has precedents in behavioral testing. CheckList organizes tests beyond aggregate accuracy \cite{ribeiro2020checklist}, AdaTest pairs model-suggested tests with human feedback \cite{ribeiro2022adatest}, and EvalLM evaluates prompts on user-defined criteria \cite{kim2024evallm}. SPADE and EvalGen generate assertions for LLM pipeline outputs and align them with human judgments \cite{shankar2024spade,shankar2024evalgen}. Our harness also generates assertions from a description of the target behavior and keeps them in a versioned library, but applies them to whole simulated conversations as diagnostic profiles within a commissioned evaluation, with grounded simulation and repeated paired comparison. Because these judgments decide both what is investigated and whether a modification is accepted, we measure repeatedly to expose their variation rather than treat a single judgment as ground truth.

Once a behavior has been identified, the remaining problem is which parts of a multi-agent system should change. Work on failure attribution asks how responsibility can be assigned to components or execution steps \cite{cemri2025why,zhang2025which}. Once the responsible components are known, prompt optimization, autonomous research agents, and reflection-based methods can improve them against a supplied objective \cite{zhou2023large,yang2023large,khattab2023dspy,yuksekgonul2024textgrad,lu2024aiscientist,yamada2025aiscientistv2,madaan2023self,shinn2023reflexion,huang2023large}. In our framework, the generated assertions define that objective and guide the \emph{Improvement Orchestrator} in proposing modifications. Harness-evolution methods similarly revise agent components, but using the same examples for improvement and assessment can overstate progress \cite{zhang2026selfharness,wang2026harnessevolution}. Our harness is revised only after an investigation ends, with human approval and without changing active or recorded decisions.

\section{System Under Test}
\label{sec:system}

\begin{figure}[!t]
  \centering
  \includegraphics[width=1.0\linewidth]{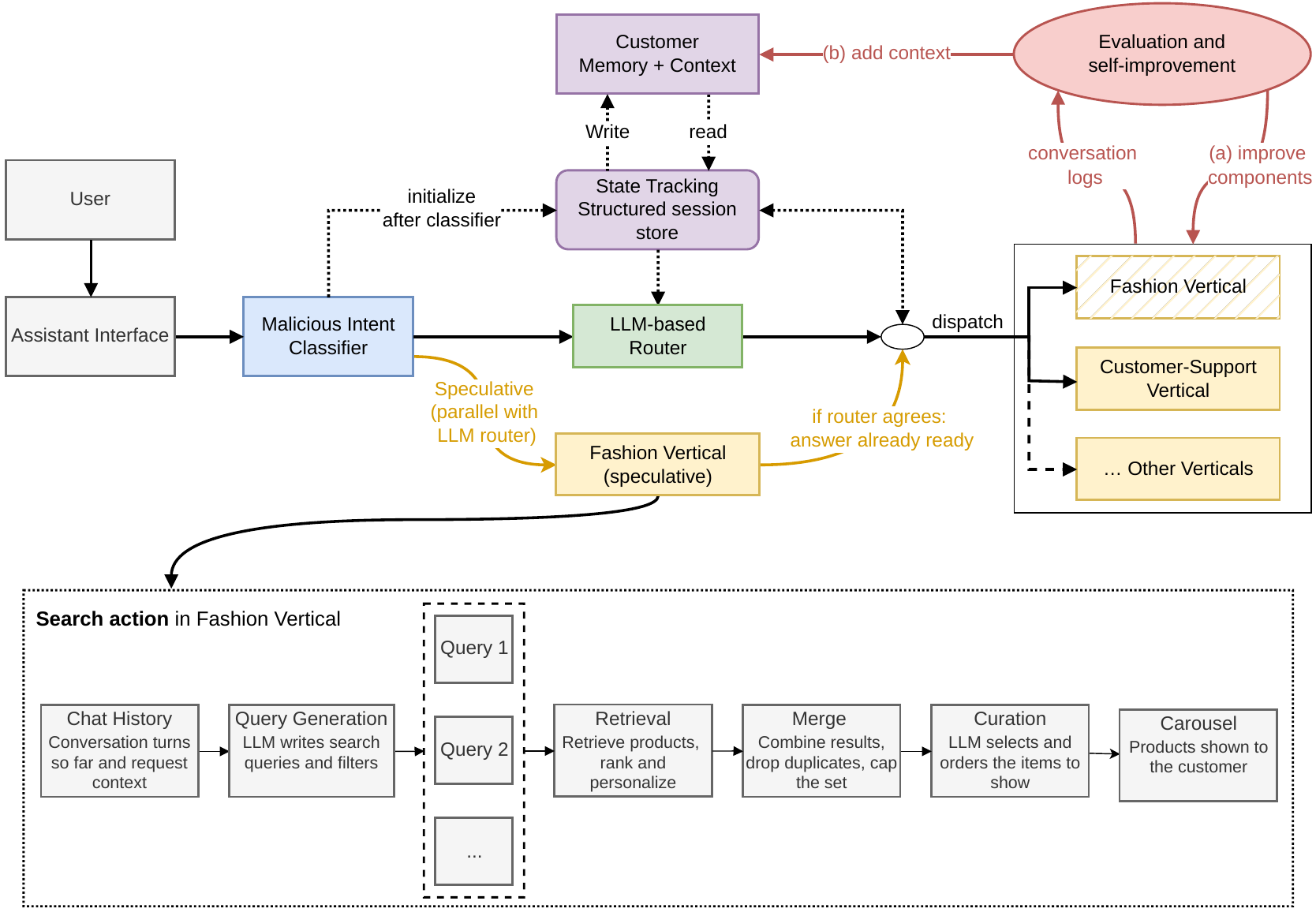}
  \caption{The \emph{Production Assistant} and its product-search action. In the upper portion, a customer request enters through the assistant interface and an LLM-based router selects the domain vertical that handles it. Conversation history, search state, and a consent-gated customer memory supply context to the verticals. The fashion vertical may start speculatively while routing is still in progress. The evaluation and self-improvement process at the upper right runs outside the serving path. The lower portion expands the search action used as the running example, from query generation through retrieval, ranking, and personalization to LLM-based curation of the carousel.}
  \label{fig:system}
\end{figure}

The \emph{Production Assistant} is a conversational shopping assistant embedded in an e-commerce platform. Customers reach it from the home page, from a product page, where the product they are viewing becomes the anchor of the conversation, and from catalog and search pages, where the current query and filters are passed along. It answers in the language of the customer's market, serves more than twenty European markets, and handles millions of customer conversations each month. Its purpose is to help customers find products, decide between them, and resolve questions about orders, in one conversation rather than across several pages.

Analyses of production conversations from four months of 2026, classified by an LLM and checked against human annotation of a sample, show the same coarse picture in every month. Close to half of all conversations are product searches, from a first request through refinements by attribute, price, or brand. Requests for style advice, outfit ideas, and pairing are the second largest group, followed by questions about a product's details, colors, sizes, and fit. Roughly one conversation in ten concerns an order, most often its status, a return, or a payment. Within this stable picture, the topics change with the season. Footwear and summer outfits rise in spring, color and requests for similar items rise in summer, and the share of style advice is highest in spring. The customer's requirements come mainly from their own message and, on product pages, from the anchor product. The attributes customers state are led by category, followed by color, brand, gender, size, and material, while style, what an item pairs with, and occasion are the most common subjective criteria. Messages range from a single word to detailed body measurements. The behaviors studied in Sect.~\ref{sec:usecases} concern category, the most frequently stated constraint, and the gender of the shopping context.

A turn passes through several LLM-based components, shown in the upper portion of Fig.~\ref{fig:system}. Two guards protect the assistant. A classifier trained in-house decides before any processing whether a message is malicious or unrelated to shopping, and blocks it if so. A general content moderation model runs while the answer is being generated and can stop it. An LLM-based router then assigns the request to a vertical. The fashion vertical is an agent that runs inside the assistant service itself and handles product search, beauty and kids' products, size and fit questions, cart actions, and small talk. Customer care is a separate service that detects the customer's language, classifies the request into cases such as a missing parcel or a return, retrieves help-center content by retrieval-augmented generation \cite{lewis2020rag}, and queries order, return, and shipment systems. Two further verticals need no LLM. Gift card requests receive a localized deep link, and messages that disclose sensitive personal data receive a refusal. Because most requests go to the fashion vertical, it starts speculatively while the router is still running, and its output is discarded only when the router selects another vertical with high confidence. This lowers latency at the cost of wasted work.

Malicious traffic is a tiny fraction of conversations, but it exists and it is deliberate. Such attempts appear regularly. Some messages instruct the assistant to ignore all previous instructions and answer an unrelated question, or to adopt a different role with a prescribed output format. Some are written in a language other than the market's to test whether the guard depends on the language. Others ask the assistant to print its system prompt, name its model and provider, or list its tools, probe for the delimiters the prompt uses, or claim to be an internal tester who needs the guardrails listed. Some multi-turn probes begin with an ordinary product request and then ask about the assistant's tools, model, and infrastructure one question at a time. None of these attempts appear to have succeeded. Both guards are components of the system under test, and an assertion that a class of message must be refused can be commissioned like any other.

Conversation state is distributed rather than held by a single component. The message history records the customer's turns, the assistant's answers, and the products shown, and a structured search state carries the parsed query, filters, and candidate lists within a turn. With the customer's consent, a memory service supplies a profile of sizes, brands, materials, and past interactions, which the fashion agent receives as part of its prompt. The shopping context, for example men's or women's fashion, arrives with the request. For questions that depend on the world outside the catalog, such as the weather at a destination or current trends, the agent can delegate to web-search sub-agents whose answers pass through moderation before they are used. Responses are streamed and can contain text, a product carousel of up to eight items, an order card, or a comparison, together with up to three suggested follow-up questions. Customers can also send an image, which a visual retrieval path matches against the catalog. Which model serves each component, and whether an optional component is active, is set through experiment configuration, so the system that serves a given customer is a configuration rather than a fixed program.

The lower portion of Fig.~\ref{fig:system} expands the search action within the fashion vertical, one of the main components examined in the reported investigations. The conversation history up to the current turn and the request context are passed to an LLM that generates one to three product queries together with structured filters for shopping context, price, sale status, sizes, and terms to exclude. Each query retrieves, ranks, and personalizes products from the current assortment. The results are combined and deduplicated against products already shown in the conversation before an LLM-based curation step selects and orders the products shown in the product carousel. When curation returns fewer products than the carousel requires, the remaining positions are filled from the pre-curation ranking without a further check. A visible problem in the carousel can originate in any of these components, not only in the one that presents the result.

Four properties of this system shaped the framework. First, a single turn involves many LLM calls across guard, router, query generation, curation, and response, each configurable by experiment, so a behavior cannot be attributed to one component from the outside. This is why the harness runs a local instance, and why every commission records which build of the assistant it evaluated. Second, at millions of conversations a month, individual conversations cannot be reviewed by people. Problems are found by automated assessments over a daily sample, and cohorts are drawn from large pools of conversations by explicit sampling strategies, based on embeddings or on customer intent. Third, the range of requests, from a first product search to a return, means that no single benchmark represents the system. Each investigation therefore commissions its own evaluation for one behavior. Fourth, the assortment, prices, stock, and personalization signals differ between markets and change daily, so the same request produces different results on different days. This is the run-to-run variation that the stored baseline and the paired repeated runs are designed to separate from the effect of a modification.

\section{Commissioning the Evaluation}
\label{sec:commission}

The \emph{Evaluation Harness} first attempts to reproduce the selected behavior in an unchanged local version of the \emph{Production Assistant}. It starts from a relevant production conversation when one exists, and otherwise from scenarios constructed from the behavior description. The objective is to reproduce the behavior, not the exact historical response. Reproduction succeeds when the reported behavior appears in the local sessions, that is, when the targeted assertions fail on enough of the cohort to compare system versions. When source conversations exist, the local failure rate should also be similar to the rate reported for them. Repeated runs then establish the baseline, which completes the commission. A commission consists of the cohort and its sampling strategy, the frozen assertions and evidence fields, the decision thresholds, and the stored baseline. If reproduction fails, the investigation stops, since we cannot optimize against a behavior that cannot be studied locally.

\subsection{Cohort Construction}
\label{sec:cohort}

The cohort is the fixed set of customer scenarios on which the unchanged and modified systems are compared. A scenario is the starting point of one simulated conversation, a first customer message or goal together with its customer context. When the target behavior is found in production logs, the \emph{Evaluation Harness} creates a pool of conversations from a defined time period and retains those that can be reconstructed for simulation. To reconstruct a conversation, the harness needs the customer's messages, the context of the request, and the products that were shown. The request context includes the page and the shopping context. The products shown are needed to judge the targeted assertions. One of the sampling strategies described below then draws the cohort from this pool.

Production data enters this process under our data protection rules. Customer identifiers are pseudonymized, scenarios carry a pseudonym of their source conversation, and this reference is excluded from every exported artifact. The judge and the \emph{Improvement Orchestrator} evaluate simulated sessions, not production conversations.

Random sampling provides a reference cohort with no imposed structure. The other strategies work either on an embedding of each conversation or on its intent. For embedding-based sampling, each conversation is embedded with a representation that weights customer messages more heavily than assistant messages, so that similarity reflects what customers asked for rather than how the assistant answered. Farthest-point sampling selects conversations that are far apart in the embedding space to show how a behavior appears across the data. Nearest-neighbor sampling finds conversations similar to known instances of the behavior. Stratified sampling groups the pool into clusters in the embedding space and selects conversations near the center of each cluster. Hybrid sampling combines a stratified core with distant conversations chosen by farthest-point sampling, so that the cohort covers both typical and unusual cases. For intent-based sampling, each conversation is assigned a customer intent (Sect.~\ref{sec:findings}), and the cohort is stratified by intent instead of by embedding cluster. The chosen strategy is recorded with the commission because it determines what the results can show. Except under random sampling, the cohort is not a random sample of production traffic. Its failure rate describes the behavior on the cohort and does not estimate how often the behavior occurs in production.

\subsection{Behavioral Assertions}
\label{sec:assertions}

Once a behavior has been selected, an LLM-based compiler translates it into explicit pass or fail conditions, which are stored in a versioned assertion library. An LLM judge later applies each condition to every simulated interaction, records the supporting evidence and reasoning, and returns a verdict. A human reviews the generated assertions before they are frozen, revising or adding assertions where needed.

An investigation typically uses several related assertions, some examining the target behavior from different perspectives and others, which we call guardrail assertions, protecting behavior that should remain unchanged. For a product carousel, for example, we evaluate the same requirement for any product, for the first product, for at least 75\% of the products, and for all products. The resulting profile indicates whether a problem affects the complete result set or only particular positions.

The judge receives the information required to decide each assertion. This evidence can include the simulated conversation, structured product data, and product images when visual assessment is required. Both the assertion text and the evidence fields shown to the judge are fixed before any modified version is evaluated. A later correction creates a new evaluation and baseline rather than changing the meaning of scores already recorded.

\subsection{Grounded User Simulation}
\label{sec:simulator}

The harness supports two modes of simulating the customer. Conversation-grounded simulation sends the first recorded customer message unchanged and uses the subsequent messages as examples of how that customer communicates and pursues the task, not as a script to reproduce. The simulator then generates each new turn in response to the system version under evaluation instead of replaying the logged continuation. Each session allows the simulated customer at most a fixed number of turns. This limit is derived from the length of the recorded conversation and kept within a lower and an upper bound, and a session ends when the simulated customer stops or the limit is reached.

When no suitable production conversation exists, goal-based simulation begins from a description of what the customer wants to achieve. Its starting conditions are the goal description, the chosen customer context, with or without personalization as the reported behavior requires, and a fixed turn limit. In both modes, the starting conditions are held fixed when the unchanged and modified systems are compared.

A separate LLM scorer rates the simulated customer on goal pursuit, task adherence, interaction efficiency, and stopping behavior, using the recorded customer messages as a soft reference when available. We validated the simulator on production-grounded cohorts spanning several languages, customer contexts, and interaction settings. On each cohort, the simulated sessions reproduced the failure that had been reported for its source conversations. We also read the simulated conversations to check that the simulated customer behaved plausibly.

\subsection{Paired Decision Rule}
\label{sec:decide}

We run the unchanged local system at least three times over the cohort to create the stored baseline, then run the modified version the same number of times over the same scenarios. Every scenario therefore has a set of baseline sessions and a set of modified sessions. For each scenario and each assertion, we average the scores within each set and subtract the baseline average from the modified average. This gives one paired difference per scenario. Scenarios that lack a score in any run, for example because a session did not complete, are excluded. The scenario is the unit of analysis. Repeated sessions make its mean more precise but do not count as independent observations.

To estimate uncertainty, the harness resamples the scenario-level paired differences with replacement and recomputes their mean \cite{efron1979bootstrap}. The recorded procedure uses 2,000 percentile bootstrap resamples with a fixed random seed. An interval entirely above zero produces \texttt{signal+}, while one entirely below zero produces \texttt{signal-}. An interval that contains zero is labeled \texttt{noise} when its half-width does not exceed a threshold fixed before any modification is tested, and \texttt{borderline} otherwise. The \texttt{noise} label means the interval is narrow on that scale. It does not mean that no difference exists. The threshold is \(0.05\) on the \(0\)--\(1\) scale and \(0.25\) on the \(1\)--\(5\) scale.

A \texttt{borderline} result calls for additional repeated runs rather than acceptance based on a positive mean. Every scored dimension, whether a pass or fail assertion or a $1$--$5$ quality score, is assigned one of three roles when the commission is frozen. A target is the behavior the modification is meant to improve, a guardrail is a behavior that must not regress, and a diagnostic, such as latency, is reported and does not enter the decision. A modified version is recommended for advancement only when a target yields \texttt{signal+}, no guardrail yields \texttt{signal-}, and a separate second evaluation satisfies the same conditions. In the gender investigation reported in Sect.~\ref{sec:usecases}, the only one of the two that used the paired rule, the target was an assertion and the task-completion score served as a guardrail. A modification that lowered it would have been rejected. Under \emph{h2} the score rose, which Table~\ref{tab:results} reports, but a higher guardrail score is not by itself a reason to accept a modification. Promising results are evaluated again before human review and possible advancement to an online A/B test.

These repeated runs are also where the cost of an investigation accrues. Judging dominates, almost entirely through input tokens, since the judge reads the transcript and the product data while its outputs are a few hundred tokens. In our current configuration a three-turn session costs roughly 4,000 tokens for the simulated customer, 13,000 for the task-completion score, and 6,000 per assertion. The category investigation reported later in Sect.~\ref{sec:usecases}, with 300 sessions and four assertions, therefore cost on the order of ten million tokens. The orchestrator also calls an LLM when it reads the traces, forms a hypothesis, and implements the modification. These calls were not metered. Since every hypothesis also triggers its own evaluation runs, the total cost of an investigation grows with the number of hypotheses tried, which the autonomy budget caps.

\section{Improvement and Harness Revision}
\label{sec:loops}

\subsection{Hypothesis-Driven Improvement}
\label{sec:hypotheses}

Once the behavior has been reproduced and the baseline established, the \emph{Improvement Orchestrator} takes the targeted assertion profile as its objective. It first examines the current system and the interaction traces to identify plausible failure mechanisms across component boundaries. We instruct it to prefer diverse hypotheses over repeated variations of the same local change. Modifications may touch prompts, tool descriptions, configuration, or code of the local system. A human sets an autonomy budget, the number of modified versions that may be compared with the baseline before the investigation returns for review. Human reviewers may contribute hypotheses, which are evaluated by the same procedure.

Each hypothesis is implemented as an isolated modification to the same clean local version of the \emph{Production Assistant}. Our current implementation evaluates hypotheses sequentially. Every modification is completed before the local system starts, and no files change during simulation.

The modified system is scored with the frozen assertions, and the detailed results determine the orchestrator's next step. If the results support the proposed mechanism, the orchestrator refines it. If they are \texttt{borderline}, it extends both versions to five runs, and if the verdict remains \texttt{borderline} or turns negative, it moves to a different explanation. The exact modification and its evaluation results are recorded before the local system is restored to its unchanged state. Failed and abandoned hypotheses remain in the investigation record for later proposals to build on.

A modification whose target assertion yields \texttt{signal+} and whose guardrails show no \texttt{signal-} is evaluated a second time with the same number of runs. If the second evaluation satisfies the same conditions, the loop stops and presents the modification for human review and a possible online A/B test. The loop also stops when the budget is exhausted or no plausible explanations remain. The loop stops at the first reproducible improvement by design. If it kept generating changes against the same frozen cohort and assertions, it would begin to optimize for that particular evaluation rather than for the behavior the evaluation stands for. Exploring further therefore requires new evidence or a new commission with a fresh cohort.

\subsection{Post-Investigation Harness Revision}
\label{sec:reflection}

The targeted evaluation stays unchanged while an investigation is active. Afterward, a reflection step reviews the investigation, both its outcome and how it was conducted. When it found no improvement or a smaller one than expected, the cause may lie in the \emph{Production Assistant}, which offered no simple fix, in the hypotheses, which were poorly chosen, or in the evaluation, which could not detect it. When the evaluation itself is responsible, reflection may propose a revised approach to selecting scenarios, simulating customers, or interpreting results. Any such revision requires human approval, creates a new version of the \emph{Evaluation Harness}, and applies only to future investigations.

In our production investigations so far, this process has produced several harness revisions. Our first comparisons pooled all sessions of a run. Repeated runs of the same scenario did not agree with one another, and the pooled interval treated the variation between scenarios as uncertainty about the modification. Small improvements were lost in this way. We therefore adopted the paired scenario-level intervals of Sect.~\ref{sec:decide}. The mismatch between embedding-based sampling and customer intent led to a separate intent vocabulary. Audits of the judge changed the information it receives and its model configuration, and corrected the noise estimate of a separate traffic-weighted benchmark report.

\section{Discussion and Lessons Learned}
\label{sec:usecases}
\label{sec:findings}

Table~\ref{tab:results} reports two investigations, one before and one after we adopted the paired decision rule. In the first investigation, production conversations showed carousels containing products outside the requested category. The cohort held 51 human-reviewed production conversations, and 50 of them were scored in every run. Four assertions evaluated the same requirement at increasing strictness, that the requested category holds for at least one product, for the first product, for at least 75\% of the products, and for all products in the carousel. The failure was addressed at the curation step of Fig.~\ref{fig:system}, where an LLM selects the products for the carousel. Its prompt already asked for products from the requested category. The hypothesis turned this preference into a strict rule that no product outside the requested category may be returned. The rule remained a prompt instruction, not a filter enforced in code. This modification was deployed on the basis of the run averages in Table~\ref{tab:results}, before we adopted the paired rule. The baseline values in Table~\ref{tab:results} describe this failure on this cohort. The first product matched the requested category in 83\% of sessions, and every product did so in 64\%, so about one carousel in three contained at least one product from another category. Because the cohort was selected to show the failure, these values do not measure the overall quality of the assistant, are not comparable to the quality metrics we monitor in production, and do not estimate how often the failure occurs in production.

In the second investigation, gender-silent requests from a men's shopping context returned products from the women's assortment. Because no suitable production conversations existed, the harness generated ten goal-based scenarios, nine gender-silent requests under a men's context plus one women's-context guardrail. Hypothesis \emph{h1} revised the tool description at the component where the men's context was being overridden. Hypothesis \emph{h2} instead made the shopping context a binding default in the system prompt of a different agent. Table~\ref{tab:results} reports five runs per version. The \emph{h2} improvement reproduced in three further runs, and the guardrail assertion passed in every run.

\medskip

These two investigations illustrate the workflow on our production system. The rest of this section turns to lessons from these and other investigations that we believe hold for production conversational agents in general.

\begin{table}[!t]
\caption{Results from the two example investigations. Each assertion score is the cohort mean pass rate for whether the requested category or shopping context holds at the stated carousel positions. Task completion, in the last row, is not an assertion but a $1$--$5$ quality score the judge assigns to the whole session. The category case used 51 production-grounded scenarios, 50 of which were scored in every run, with three runs per version. It predates the paired-bootstrap rule, so its values are means and standard deviations across runs, and dashes mark the unrecorded intervals and verdicts. The gender rows cover the nine gender-silent scenarios over the initial five runs per version, and intervals are paired 95\% percentile bootstrap intervals over scenario-level differences. They are wide because the cohort provides only nine paired differences, one per scenario. Cohort size is set per behavior, and an inconclusive verdict triggers additional runs rather than acceptance. Pass rates refer to the stated behavior on a cohort selected to show the failure. They do not measure overall assistant quality or how often the behavior occurs in production.}
\label{tab:results}
\centering
\scriptsize
\setlength{\tabcolsep}{2.5pt}
\begin{tabular}{@{}llccrcl@{}}
\hline
Case & Assertion or score & Baseline & Modified & $\Delta$ & 95\% CI & Verdict \\
\hline
Category & Any product & $0.913 \pm 0.012$ & $0.947 \pm 0.023$ & $+0.033$ & -- & -- \\
 & First product & $0.833 \pm 0.023$ & $0.893 \pm 0.031$ & $+0.060$ & -- & -- \\
 & $\geq$75\% of products & $0.733 \pm 0.076$ & $0.747 \pm 0.031$ & $+0.013$ & -- & -- \\
 & All products & $0.640 \pm 0.040$ & $0.647 \pm 0.041$ & $+0.007$ & -- & -- \\
Gender, \emph{h1} & First product & $0.733$ & $0.800$ & $+0.067$ & $[-0.022,+0.178]$ & \texttt{borderline} \\
Gender, \emph{h2} & First product & $0.733$ & $1.000$ & $+0.267$ & $[+0.067,+0.511]$ & \texttt{signal+} \\
 & Task completion & $4.27$ & $4.96$ & $+0.69$ & $[+0.22,+1.20]$ & \texttt{signal+} \\
\hline
\end{tabular}
\end{table}

\paragraph{Visible behavior does not identify the right intervention.} Although unsurprising, this observation determined the course of most investigations. A failure may be introduced early in the pipeline and become visible only in the carousel. Both improvements in Table~\ref{tab:results} were short prompt changes made where the constraint is set, not where the failure showed up. In the gender case, \emph{h1} modified the component where the failure was first suspected and was \texttt{borderline}. \emph{h2} expressed the same rule in another agent and produced a clear improvement. We instruct the \emph{Improvement Orchestrator} to propose hypotheses along the complete interaction path rather than repeatedly modify the component where the failure is visible. This is also why we run the system under test locally. A local instance lets us observe and evaluate every component along the interaction path, which requires an evaluation suite that reaches into the individual components of the multi-agent system. The harness can propose such component-level assertions itself when an investigation needs them.

\paragraph{Targeted assertions connect evaluation with improvement.} Assertions that tie a specific behavior to the components that produce it make an evaluation actionable. General measures such as relevance, intent understanding, recommendation quality, and customer satisfaction help find low-quality production interactions, but they provide little indication of what should change. Commissioned evaluation replaces these broad measures with assertions generated for the selected behavior, which examine it from several angles and at several levels of strictness. In Table~\ref{tab:results}, the category modification raised the first-product and any-product assertions, while the 75\% and all-products assertions changed little. This matches a mechanism that affects prominent positions rather than the whole result set. These profiles give the \emph{Improvement Orchestrator} a concrete basis for hypotheses, and afterward a check of whether the modification affected the assertions that the hypothesis predicted.

\paragraph{Run-to-run variation must be separated from differences between scenarios.} Two kinds of variation appear in the sessions. The first is variation between scenarios, because some scenarios are harder than others. The second is variation between repeated runs of the same scenario, because the assistant, the simulator, and the judge are stochastic. Only the second kind shrinks with more runs. If all sessions are pooled, the difference in difficulty between scenarios is counted as uncertainty, and a small improvement that holds in every scenario is lost in it. We therefore compare each scenario with itself. Averaging the runs of a scenario reduces the second kind of variation, and comparing the two versions on the same scenario removes the first.

\paragraph{Diversity depends on how conversations are represented.} In our analysis of production traffic and in cohort construction, geometric distance in the embedding space reflected vocabulary, product category, and language more readily than customer intent, so two conversations can be distant while exercising the same behavior or close while expressing different needs. We addressed this with a separate intent vocabulary of ten customer intents with about forty subtopics. Each conversation in the pool is assigned an intent, and cohorts can be stratified by intent rather than by embedding cluster, since one intent typically spans several regions of the embedding space.

\paragraph{The judge must be tested as software.} The judge can decide an assertion only if it receives the information the assertion needs. One assertion could not assess its target behavior until the relevant result information was added to its evidence. A second defect concerned configuration. The judge temperature was set to zero, but the client did not pass this setting to the model, so it had no effect. A third concerned statistics. A separate benchmark report used differences between scenarios where it should have measured variation across repeated runs of the same scenario, although this defect did not reach the paired comparison used for decisions. In each case the output appeared plausible, and the first two could have changed a decision. Judge inputs, the executed configuration, and the statistical procedure each require direct tests.

\paragraph{A frozen evaluation can become outdated.} A commission freezes the cohort, the assertions, the evidence shown to the judge, and the decision thresholds, so that every modified version is judged by the same criteria. Shared components, the available products, or the behavior represented by the scenarios may change afterward, and the stored baseline then no longer describes the current system. Each commission should record the assistant build and judge models it depends on, and its baseline should be rerun when either changes.

\paragraph{The number of hypotheses is not fixed in advance.} The \emph{Improvement Orchestrator} compares one modified version after another with the same stored baseline and stops at the first reproducible improvement or when the budget is exhausted, so the number of comparisons is decided by the results themselves. Each comparison carries a small probability of a false \texttt{signal+}, and over a loop these probabilities compound. With independent comparisons at a nominal level of 0.05, the probability of at least one false \texttt{signal+} among fifty null hypotheses exceeds 0.9. We reduce this risk by fixing target and guardrail roles before testing, requiring a second evaluation before advancement, and capping the number of hypotheses. These steps lower the risk but do not bound it.

Classical corrections for multiple testing do not fit this setting. They divide the error budget among a number of tests fixed before the data are seen, and the loop has no such number. Using the running count of hypotheses instead makes the threshold depend on when a modification is tried. A Bonferroni correction would demand a nominal level of 0.001 for the fiftieth hypothesis and 0.0001 for the five hundredth, so the same modification would be accepted early in a loop and rejected late in it, for reasons unconnected to the modification.

The framework built for this situation is anytime-valid inference with e-values \cite{grunwald2024safe,ramdas2023game}. An e-value is a nonnegative statistic whose expectation under the null hypothesis is at most one. The product of a sequence of e-values, one per batch of evidence, forms a test martingale, and Ville's inequality, proved in 1939 \cite{ville1939}, states that such a martingale exceeds $1/\alpha$ at any time with probability at most $\alpha$. This single inequality gives type I error control under optional stopping and optional continuation with one threshold for every hypothesis, however many are tested. Gr{\"u}nwald et al.\ derive the corresponding safe t-test in their Section 4.3, which can be applied to every new batch of evidence, and the e-BH procedure of Wang and Ramdas extends the guarantee to false discovery rate control across hypotheses under arbitrary dependence \cite{wang2022fdr}. The price is a minimum effect size that must be fixed before testing, a role our noise threshold already plays. Recent work applies e-processes to LLM evaluation, for certified confidence intervals under optional stopping \cite{zhou2026celeus} and for attributing drift between a system and its judge \cite{li2026drift}.

As a post hoc check, we recomputed the interval for the \emph{h2} first-product assertion with the betting method of Waudby-Smith and Ramdas \cite{waudbysmith2024betting}, which has guaranteed coverage for bounded outcomes at any sample size. The per-scenario results were not stored, so we reconstructed them from the recorded run totals. The lower end of the betting interval lies between 0.00 and 0.01, whereas the percentile bootstrap in Table~\ref{tab:results} gives 0.067. With this method, \emph{h2} would have been \texttt{borderline} at nine scenarios. Doubling the same differences to eighteen scenarios raises the lower end to 0.08, which would be \texttt{signal+}. Agents that improve themselves by testing many hypotheses are, in effect, best evaluated with an inequality from almost a century ago, and we intend to adopt it in the loop.

\section{Limitations}
\label{sec:limits}

All results reported here are offline. An offline improvement does not by itself establish gains in customer satisfaction or business outcomes in live traffic. Those require controlled A/B tests. Real customers are also represented by the simulator. We validated it by checking that it reproduces the reported failures on its source conversations and by reading the simulated conversations. We have not measured how closely the simulated customer matches real customer behavior across a wide range of intents and scenarios.

The paired bootstrap intervals of the decision rule treat the judge and the production environment as fixed. In practice, judge models differed across investigations, and the recorded scores predate the fix of the judge temperature. Differences between judge configurations are not part of the reported uncertainty.

The decision rule tests several assertions and hypotheses without a formal multiplicity correction \cite{carterette2012multiple}. Because the number of hypotheses is not fixed in advance, classical corrections do not apply cleanly, and the safeguards described in Sect.~\ref{sec:usecases} lower the risk of false discoveries but do not bound it. The anytime-valid analysis in that section was computed after the fact. Its input, the difference per scenario, was reconstructed from the recorded run totals, and the interval was computed with our own implementation of the method. Finally, with nine paired scenarios the gender case has limited power, and percentile bootstrap intervals under-cover at this size \cite{hesterberg2015bootstrap}.

\section{Conclusion}
\label{sec:conclusion}

We have described how a production conversational agent is evaluated and improved one behavior at a time. Logs cannot be replayed, since a modification changes the conversations the agent will have, and even the unchanged agent varies from run to run as its LLM components and its environment change. Instead, each investigation commissions its own evaluation, a fixed cohort of scenarios with frozen assertions, and compares every modified version with a stored baseline through grounded simulation and paired repeated runs. In the two investigations reported here, one modification was deployed and another was confirmed offline. Assertion profiles at several levels of strictness showed which part of the result a failure affected, and repeated measurement kept run-to-run fluctuations from being mistaken for improvements. Auditing the judge itself uncovered defects that plausible-looking scores had concealed, which shows that the evaluation must be tested as carefully as the system it evaluates. The harness learns from completed investigations under human approval, while the decisions of those investigations stay as recorded. We expect the same approach to apply wherever an LLM-based agent must be improved against behaviors observed in production, provided that its users can be simulated closely enough to reproduce those behaviors.

\bibliographystyle{splncs04}
\bibliography{refs}

\end{document}